\documentclass{PoS}

\title{Dilaton vs Higgs: Nearly Conformal theory with confinement-like pattern}

\ShortTitle{Dilaton vs Higgs: Nearly Conformal theory with confinement-like pattern}

\author{\speaker{Gennady Kozlov}\\
        JINR\\
        E-mail: \email{kozlov@jinr.ru}}

\abstract{We consider the model containing a dilaton vs Higgs boson in the nearly conformal sector (NCS). The potential of a dilaton in NCS is linearly rising with distances. The light scalar dilaton would be one of the best candidates to explain the LHC data in recent discovery of a Higgs-like resonance at 125 GeV.}

\FullConference{Xth Quark Confinement and the Hadron Spectrum\\
		 8-12 October 2012\\
		 TUM Campus Garching, Munich, Germany}

\begin{document}

\section{ Introduction.}  It is known that the electroweak symmetry breaking (EWSB) at the
scale $v\simeq$ 246 GeV 
can  be triggered by spontaneous breaking of  scale  symmetry at an  energy  $f\geq v$ [1,2]. In this scenario, there is a nearly conformal dynamics at a scale
$ \sim 4\pi f$ below which the scale symmetry is broken and one feeds into an electroweak (EW) sector. In
the spectrum there is an EW singlet scalar field $\chi (x)$, the dilaton mode, that is the
pseudo-Goldstone boson associated with the spontaneous breaking of conformal symmetry.
The real dilaton field $\sigma (x)$ is parametrized by $\sigma (x) = \chi (x) - f$ with the order parameter $\langle \chi (x)\rangle = f$. The Higgs-boson can be seen as a dilaton in the limit case $ f = v$.  The mass of the $\sigma$-dilaton is naturally light, $m^{2}_{\sigma} = \epsilon\cdot f^{2}$, where the small parameter $\epsilon$ controls the deviations from exact scale invariance. The dilaton becomes massless when the conformal invariance is recovered.

Since of its pseudo-Goldstone nature, the dilaton could be the  messenger field between
the Standard Model (SM) fields and the hidden sector. For example, the dilaton field itself can be lighter than, e.g., the dark matter (DM) particles, and be the dominant product of DM annihilation.
Note that if one assumes that both SM and DM are fully embedded in the conformal
sector, one can propose that the dilaton is the dominant messenger between the DM, the SM and
the unparticle stuff. The latter itself with the scaling dimension $d$ may appear as a non-integer
number $d$ of invisible particles [3].

We assume that the dilaton could be the dominant origin of the unparticle production through the SM fields. 
For this, the model in which  the decay of a dilaton into a vector unparticle $U$ and a single photon, $\sigma\rightarrow\gamma\,U$, is studied. 
In EW sector the latter decay process transforms into $\sigma\rightarrow\gamma\gamma$.
Signals of the unparticle stuff can be detected through the missing energy and momentum distribution carried away by  the unparticle. The coupling of a dilaton to the unparticle stuff is through the loop composed with the  quark fields flowing in the loop. The attractive feature of the decay  $\sigma\rightarrow\gamma\,U$ is that the photon energy has a continuous spectrum in the rest frame of $\sigma$, in contrast to, e.g., $\sigma\rightarrow\gamma\gamma$ or $\sigma\rightarrow\gamma\,Z$. The mode $\sigma\rightarrow\gamma\,U$ would predict a useful tool for study of new physics at nearly conformal sector in respective broad range of the dilaton mass up to the order $O(1~TeV)$, and the distiguishing the dilaton from the SM Higgs-boson $H$, restricted by its mass with $\sim$ 125-126 GeV in the decays like $H\rightarrow\gamma\gamma$, $H\rightarrow\gamma\,Z$,  $H\rightarrow\gamma\,U$, etc.

\section{Couplings.} The production of the dilaton can be  through the gluon-gluon fusion, $g\,g\rightarrow \sigma$.
Since the couplings $\sigma g\,g$ are crucial for collider phenomenology, it has been shown [2] that these couplings can be significantly enhanced under very mild assumption about high scale physics. At energies below the scale $4\pi f$ the effective dilaton couplings to massless gauge bosons are provided by the SM quarks  lighter than the dilaton:
$ [c_{EM} (F_{\mu\nu})^{2} + c_{s} (G^{a}_{\mu\nu})^{2}]\sigma/(8\,\pi\,f) $. Here, $F_{\mu\nu}$ and
$G^{a}_{\mu\nu}$ are the electromagnetic (EM) and gluon fields strength tensors, respectively;
$c_{EM} = - \alpha\cdot 17/9$ if $m_{W} < m_{\sigma} < m_{t}$, $c_{EM} = - \alpha\cdot 11/3$
if $m_{\sigma} > m_{t}$;
$c_{s} = \alpha_{s}\cdot (11 - 2\,n_{light}/3)$; $n_{light}$ is the number of quarks lighter than the dilaton;
$\alpha$ and $\alpha_{s}$ are EM and strong coupling constants, respectively; $m_{W}$ and $m_{t}$ are masses of the $W$-boson and the top-quark, respectively. The second term
in the effective coupling above mentioned indicates a $(33/2 - n_{light})$-factor increase of the
coupling strength compared to that of the SM Higgs boson.
The upper limit of $f$ is estimated in [4]: 
$f < 5.33$ TeV if the dilaton is lighter than the top quark, or $f < 4.87$ TeV  otherwise.

\section{ Model.}
The model is formulated in terms of a Lagrangian which features: the dilaton field $ \sigma (x)$ as the local operator and from which the vector potential $A_{\mu} (x)$ is derived, the conformal field given by the operator $O_{U}^{\mu}$ and a set of the SM fields. The conformal invariance can be broken by the couplings with non-zero mass dimension effects.
The Lagrangian density (LD) with a small explicit breaking of the conformal symmetry is $ L = L_{1} + L_{2}$, where
\begin{equation}
\label{e1}
L_{1} = -B\partial_{\mu}A^{\mu} + \frac{1}{2\xi}B^{2} -\frac{1}{\Lambda_{U}^{d-3}}(A_{\mu}
- \partial_{\mu} \sigma)O^{\mu}_{U} + \bar\psi (i\hat {\partial} - m +g\hat {A})\psi
 -  \frac {\sigma}{f}\sum_{\psi} \left (m + \epsilon y_{\psi}v\right ) \bar\psi\psi  ,
\end{equation}
\begin{equation}
\label{e2}
L_{2} =  \frac{1}{\Lambda_{U}^{d-1}}\left [\sum_{q} \bar \psi (c_{v}\,\gamma^{\mu} -  a_{v}\,\gamma^{\mu}\gamma_{5})
\psi\, O_{{U}_{\mu}} + \frac{1}{\Lambda^{2}_{U}} W^{a}_{\mu\alpha}W^{a\mu}_{\beta}\left (\partial^{\alpha}
O^{\beta}_{U} + \partial^{\beta}O^{\alpha}_{U}\right )\right ] .
\end{equation}
The field $B$ plays the role of the gauge-fixing Lagrangian multiplier, and it remains free.  We assume $\xi\neq\infty$  in (\ref{e1}) since otherwise the model becomes trivial.
The unparticle vector operator $O_{U}^{\mu}$ describes a scale-invariant hidden sector that possesses IR fixed point at a high scale $\Lambda_{U}$, presumably above the EW scale; $c_{v}$ and $a_{v}$ are vector and axial-vector couplings.

A dilaton acquires a mass and its couplings to quarks can undergo variations from the standard form. In particular, since scale symmetry is violated by operators involving quarks, shifts in the dilaton Yukawa couplings to quarks can appear. This is given in (\ref{e1}) by $\epsilon $ which parametrizes the size of the deviation from exact scale invariance [5]. In LD (\ref{e1}) the nine additional contributions to Yukawa couplings $y_{\psi}$ are taken into account ($y_{\psi}$ are $3\times 3$ diagonal matrices in the flavor space);   $\psi (x)$ stands for the spinor field with the mass $m$.  

In the model considered here, the only  SM quarks contribution is dominated, because the $W$-boson loop contribution is suppressed by two more powers of $\Lambda_{U}$ in (\ref{e2}), and due to significantly large value of $\Lambda_{U}$ one can ignore it.

The equations of motion  are ($\nabla \equiv \partial /\partial x_{\mu}$)
$$\partial_{\mu} \sigma \simeq A_{\mu} -  \frac{1}{\Lambda_{U}^{2}} \bar \psi (c_{v}\,\gamma_{\mu} -  a_{v}\,\gamma_{\mu}\gamma_{5}) \psi, \,\,\, \partial_{\mu}\,A^{\mu} = \xi^{-1}\,B, $$

$$\partial_{\mu} B = - J_{\mu} + \frac{1}{\Lambda_{U}^{d-3}}\,O_{U_{\mu}}, \,\,\,\,\,\,
J_{\mu} = g\,\bar{\psi}\,\gamma_{\mu}\,\psi,$$

$$\frac{1}{\Lambda^{d-3}_{U}}\,\partial_{\mu}O_{U}^{\mu} +  \frac{1}{f} (m + \epsilon\,y_{\psi}\,v)
\bar {\psi}\,\psi = 0, $$

$$\left [i\,\hat {\partial} - m\,\left (1+\frac{\sigma}{f}\right ) + g\,\hat {A} - \frac{\sigma}{f}\,
\epsilon\,y_{\psi}\,v + \frac{1}{\Lambda_{U}^{d-1}}\, O^{\mu}_{U} (c_{v}\,\gamma_{\mu} -  a_{v}\,\gamma_{\mu}\gamma_{5})\right ]\psi = 0. $$

In the nearly conformal sector (NCS) supported by the weakly changing operator $O_{U}^{\mu}$ in the space-time and the conservation of the current $J_{\mu}$, the $ \sigma (x)$-field  looks like the dipole field obeying the equation of the 4th order 
\begin{equation}
\label{e5}
\lim_{m_{\sigma}\rightarrow 0} \left (\nabla^{2} + m^{2}_{\sigma} \right )^{2} \sigma (x) \simeq 0 
\end{equation}
and the canonical commutation relation (see, e.g., the book [6])
$$ [\sigma (x), \sigma (y) ] =  \frac{1}{\xi}\,\int \frac{d^{4} p}{(2\,\pi)^{3}}\, sgn (p^{0})\,\delta^{\prime} 
(p^2)\,e^{-i\,p\,(x-y)} = \frac{1}{8\,\pi\,i\,\xi}\, sgn (z^{0})\,\theta (z^{2}), z = x-y, $$
where $sgn (p^{0})\,\delta^{\prime} (p^2)$ is  well-defined as the odd homogeneous generalized function from the space $S^{\prime} (\Re _{4})$ of the temperate distributions on $\Re_{4}$.

\section{Propagator.} To find the propagator of the $\sigma (x)$-field in NCS we use the two-point Wightman function (TPWF) in the form [7] 
$W(z) = \langle\Omega, \sigma (x)\,\sigma (y)\,\Omega\rangle = -i\,\xi^{-1}\,E^{-} (z)$, 
where $E^{-} (x)$ is the only distribution among the solutions of the equation 
$\left(\nabla^{2} \right )^{2} W (x) = 0$ obeying locality, Poincare covariance and the spectral conditions, however not positive definiteness of the metric.   The vacuum $\Omega$-vector satisfies the following conditions: $\sigma ^{-} (x)\vert\Omega\rangle = 0$, $\langle\Omega,\Omega\rangle = 1$, where $[\sigma^{-} (x)]^{*} = \sigma^{+} (x)$ in the decomposition $\sigma (x) = \sigma^{-} (x) + \sigma^{+} (x)$.
 The solutions of Eq. (\ref{e5}) can be classified by their TPWF's. One has  
\begin{equation}
\label{e7}
  E^{-} (x) = \int \frac{d^{4} p}{(2\,\pi)^{3}}\, \theta (p^{0})\,\delta^{\prime} 
(p^2)\,e^{-i\,p\,x} =  \frac{-1}{(4\,\pi)^{2}}\,\ln (-\mu^{2}\,x^{2} + i\,\epsilon\,x^{0}) 
 \end{equation}
which is the negative-frequency part of the generalized function (distribution) in $E(x) = E^{+} (x) + E^{-} (x) = (8\,\pi)^{-1}\, \theta (x^{2})\,sgn (x^{0})$; $\mu$ is the positive constant required for dimensioneless reasons.
To separate the IR parameter $\mu$-dependence, the TPWF (\ref{e7}) can also be given in the form\\  
$(-4\,\pi)^{-2}\left\{\ln \vert \mu^{2}\,x^{2}\vert + i\,\pi\,sgn (x^{0})\,\theta (x^{2})\right\} $.

The Fourier transform of $E^{-} (x)$ is $\tilde E^{-} (p) = 2\,\pi\,\theta (p^{0})\,\delta ^{(1)} (p^{2}, \tilde\mu^{2})$, where $\tilde\mu = 2\,e^{-\gamma +1/2}\,\mu$, $\gamma = - \Gamma ^{\prime} (1)$ being the Euler's constant. The functional $  \delta ^{(1)} (p^{2}, \tilde\mu^{2})$ is defined on the space $S (\Re _{4})$ of the complex Schwartz test functions on $\Re _{4}$ as [8] 
$$\delta ^{(1)} (p^{2}, \tilde\mu^{2}) = \frac{1}{16} \left (\frac{\partial ^{2}}{\partial p^{2}}\right )^{2} \left [\theta (p^{2})\,\ln\frac{ p^{2}}{\tilde\mu ^{2}}\right ]. $$
The presence of the parameter $\mu$ in $E^{-} (x)$ (\ref{e7}) breaks its covariance under dilatation transformations $x_{\mu}\rightarrow \lambda\,x_{\mu}$ ($\lambda > 0$) and implies spontaneously symmetry breaking of the dilatation invariance of (\ref {e5}). This is one of the reasons for the  special role of the dipole field $\sigma (x)$ in what follows.

The Green's function in $\Re ^{4}$ space-time is given by 
$G (z) = \langle\Omega, T[\sigma (x)\,\sigma (y)]\Omega\rangle = -i\,\xi^{-1}\,E_{c} (z), $
where the causal function 
$$  E_{c} (x) = \theta (x^{0})\,E^{-} (x) + \theta (-x^{0})\,E^{+} (x) = \frac{1}{i\,(4\,\pi)^{2}}\,\ln (-\mu^{2}\,x^{2} + i\,\epsilon) $$
satisfies the following equations
$$ \nabla ^{2} E_{c} (x) = \frac {i}{4\,\pi^{2}}\,\frac{1}{-x^{2}_{\mu} + i\,\epsilon}, \,\, 
(\nabla ^{2})^{2} E_{c} (x) = \delta ^{4} (x). $$

In $\Re _{4}$-momentum space the propagator is given in terms of distributions 
$$\tilde G (p) = \frac{-1}{(4\,\pi)^{2}\,\xi}\,\int d^{4} x\,e^{i\,p\,x}\,\ln (-\mu^{2}\,x^{2} + i\,\epsilon). $$
One can calculate $\tilde G (p)$ through $\tilde G (p) = (\partial ^{2}/\partial p^{2}) H(p)$, where [7]
$$ H(p) = \frac{-1}{(4\,\pi)^{2}\,\xi}\,\int d^{4} x\,e^{i\,p\,x}\,
\frac{\ln (-\mu^{2}\,x^{2} + i\,\epsilon)}{-x^{2}_{\mu} +i\epsilon} . $$
Finally, the result is 
$$H (p) = \frac{i}{4\,\xi}\,\frac{\ln\left [e^{2\gamma} (-p^{2} -i\,\epsilon)/(4\,\mu^{2})\right ]}{ p^{2} + i\,\epsilon}, $$
which leads to 
\begin{eqnarray}
\label{e60}
\tilde G(p) = \frac{1}{2\,i\,\xi}\,\frac{\partial}{\partial p^{\mu}}\left \{\frac{ p^{\mu}\,\left [\ln\left ( -p^{2}/\tilde\mu^{2} - i\,\epsilon\right )\right ]}{ (p^{2} + i\,\epsilon)^{2}}\right \}.
\end{eqnarray}
The following equation is straightforward: $(-p^{2})^{2}\, i\,\xi\,\tilde G(p^{2}) =1 $. 
The differentiation over $p_{\mu}$ in (\ref{e60}) with $\partial /\partial p^{\mu}$ being the weak derivative has to be understood in the sense of distribution where for any test function $u(p)$ we have 
$$ \int \tilde G(p)\,u(p)\,d^{4} p = \frac{i}{2\,\xi} \int d^{4} p \,\frac{\ln (-p^{2}/\tilde\mu^{2} - i\,\epsilon)}{(p^{2} + i\,\epsilon)^{2}}\, p^{\mu}\,\frac{\partial}{\partial p^{\mu}} u(p)$$
and the extra power of momentum $p^{\mu}$  explicitly eliminates IR divergence.

The lowest order (potential) energy of a static "charge" is given by the  Fourier transform  $(\vert \vec x \vert \equiv r)$
$$\varepsilon (r) \sim\int \frac{d^{3}\vec p}{(2\,\pi)^{3}}\,e^{i\vec p\vec x} \,D (p^{0} =0, \vec p; M)$$
with $D(p, M) =M^{2}\,\tilde G(p)$, $M$ has the dimension one in mass units. Using the propagator $\tilde G (p)$ in the form 
$$\tilde G(p) = \frac{1}{4\,\xi\,i}\,\frac{\partial ^{2}}{\partial p^{2}}
\left\{\frac{\ln\left [e^{2\,\gamma}\, ( -p^{2}/\tilde\mu^{2} - i\,\epsilon )\right ]}{ -p^{2} - i\,\epsilon}\right \} $$
which is equivalent to (\ref{e60}), one can find 
$$\varepsilon (r) \sim \frac{M^{2}}{8\,\pi\,\xi}\, r\,\left [a + b\ln(r\tilde\mu)\right ],$$
where $a$ and $b$ are the constants. Thus, the energy of a dilaton in NCS is linearly rising as $r$. The result is stable both at short and large distances in any finite order of perturbation theory.

The dominant effective potential for heavy quark and antiquark bound states at small distances is (see for details [9])
$$V_{eff}(r)\sim -\frac{C_{F}}{r}\,\alpha_{s} (m_{q}) -\frac{\lambda (m_{q}, \eta_{\sigma q})}{r}\,\exp({-m_{\sigma} \,r})$$
with
$$\lambda (m_{q}, \eta_{\sigma q}) =\frac{m^{2}_{q}}{4\,\pi\,f^{2}}\,\eta^{2}_{\sigma q},$$
where $\eta_{\sigma q}$ reflects the model "flavor" in the strength of the interaction between the dilaton $\sigma$ and heavy quarks $q$ ($\eta_{\sigma q} =1$ in SM, otherwise, $\eta_{\sigma q} >1$), $C_{F} =4/3$ for SU(3) group. The lower bound on heavy quark mass $m_{q}$ is given as 
$$m_{q} > \frac{f}{\eta _{\sigma q}}\,\left (4\,\pi\,C_{F}\,\alpha_s\right )^{1/2} $$
which can exceed the top quark mass even if $f\simeq v$ and $\eta_{\sigma q} = 1 +\delta$ as $\delta < 1$.

\section{Decay rate.}
The Lorentz invariant matrix element of the decay $\sigma\rightarrow\gamma U$ is 
$ M_{\sigma\gamma U} = \epsilon^{\star}_{\mu}(k,\lambda)\, \epsilon^{\star}_{\nu}(P_{U},\lambda_{U})\,M^{\mu\nu}_{\sigma\gamma U}$, where $\epsilon_{\mu}(k,\lambda)$ and $\epsilon_{\nu}(P_{U},\lambda_{U})$ are the wave functions of the photon (with momentum $k$ and the polarization $\lambda$) and the unparticle (with momentum $P_{U}$ and the polarization $\lambda_{U}$); $M^{\mu\nu}_{\sigma\gamma U} = [P^{\mu}_{U}\,k^{\nu} - g^{\mu\nu}\,(P_{U}\cdot k)]\,A + \epsilon^{\mu\nu\alpha\beta}\,k_{\alpha}\,P_{U\beta}\,B$. The amplitude $A$ is induced by the quark loop analogously as in the decays of a scalar Higgs-boson into two photons [10],   or $H\rightarrow \gamma Z$ [11], or $H\rightarrow \gamma U$ [12]. The other amplitude $B$ vanishes because of the scalar nature of the dilaton. 

In case the EWSB is triggered by spontaneous breaking of the scale symmetry at $f\geq v$, the decay amplitude $A$ is
\begin{equation}
\label{e8}
A(x_{q},y_{q}) = \frac{3\,\alpha\,v}{\pi\,m_{W}\,s_{W}\,\Lambda^{d-1}_{U}\,f}
\sum_{q} c_{v}\,e_{q}  \left [I(x_{q}, y_{q}) - J(x_{q}, y_{q})\right ]
\end{equation}
with $x_{q}= 4\, m_{q}^{2}/m^{2}_{\sigma}$, $y_{q}= 4\, m_{q}^{2}/P^{2}_{U}$
for  the quarks $q$  with the mass $m_{q}$ and the electric charge $e_{q}$;
$s_{W}\equiv \sin\theta_{W}$, $\theta_{W}$ is the angle of weak interactions. The axial-vector coupling $a_{v}$ in (\ref{e2}) does not contribute to $A(x_{q},y_{q}) $ because of charge conjugation constraint.
We deal with the following expressions for $I(x_{q}, y_{q})$ and $J(x_{q}, y_{q})$ [12]:
$$I(x_{q}, y_{q}) =\frac{x_{q}\,y_{q}}{x_{q} - y_{q}}\left\{ \frac{1}{2} -  J(x_{q}, y_{q}) +
\frac{x_{q}}{x_{q} - y_{q}}\left [g(x_{q}) - g(y_{q})\right ]\right\}, $$
$$J(x_{q}, y_{q}) = \frac{x_{q}\,y_{q}}{2 (x_{q} - y_{q})}\left [f(y_{q}) - f(x_{q})\right ].$$


For heavy quarks   ($x_{q} \geq 1$), obeying the conditions $(4\,m_{q}/m_{\sigma}) >
x_{q} \geq  (2\,m_{q}/m_{\sigma})$, we use 
$$
f(x_{q})= {\left(\sin^{-1}\sqrt {\frac{1}{x_{q}}}\right )}^{2},\,\,
g(x_{q}) = \sqrt {x_{q} - 1}\,\sin^{-1} \left (\sqrt {\frac{1}{x_{q}}}\right ) $$
and
$$ f(y_{q})= {\left(\sin^{-1}\sqrt {\frac{1}{y_{q}}}\right )}^{2},\,\,
g(y_{q}) = \sqrt {y_{q} - 1}\,\sin^{-1} \left (\sqrt {\frac{1}{y_{q}}}\right ), $$
where $1\leq y_{q} < (1-\epsilon_{\gamma})^{-1}$, $\epsilon_{\gamma} = 2\,E_{\gamma}/m_{\sigma}$.
The energy of the photon $E_{\gamma} = (m_{\sigma}^{2} - P_{U}^{2})/(2\,m_{\sigma})$ is restricted
in the window $[0, m_{\sigma}/2]$.

The energy distribution of the emitted photon in the decay width $\Gamma (\sigma\rightarrow\gamma U)$ is
$$\frac{d\Gamma (\sigma\rightarrow\gamma U)}{d E_{\gamma}} = \frac{A_{d}}{(2\,\pi)^{2}}\,m_{\sigma}\,
E^{3}_{\gamma}\left (P^{2}_{U}\right )^{d-2}\,{\vert A(x_{q},y_{q})\vert } ^{2}, $$
where [3]
$$A_{d} = \frac{16\,\pi^{5/2}}{(2\,\pi)^{2d}}\, \frac{\Gamma(d+1/2)}{\Gamma(d-1)\,\Gamma(2d)}.$$

\begin{figure}[h!]
\renewcommand{\figurename}{Fig.}
  \centering
    \includegraphics[width=\textwidth, height = 85mm]{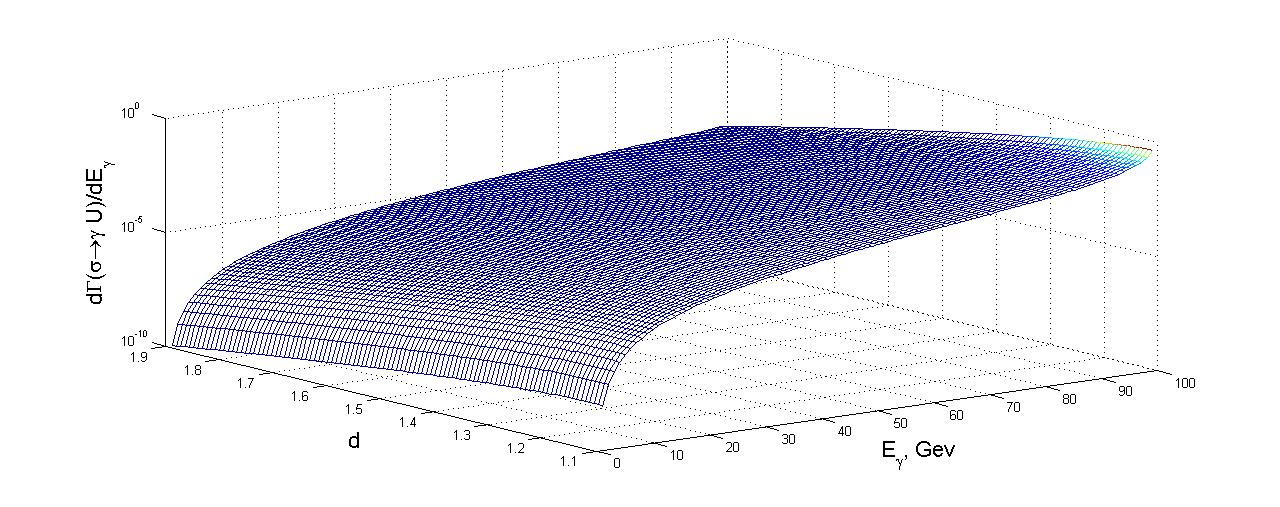}
    \caption{{ \it Energy spectrum of the photon in decay $\sigma\rightarrow\gamma U$ for various values of $d$. }}
  \label{fig:secondgraph}
\end{figure}


In Fig. 1, we show the energy spectrum of the emitted photon in decay $\sigma\rightarrow\gamma U$ for various
values of $d$ with the dilaton mass $m_{\sigma}$ = 200 GeV, $c_{v}$ = 1, $\Lambda_{U}$ = 1 TeV, $f\simeq v$. The only top quarks in the loop are included for the calculations because of the negligible contributions from lighter quarks  in the amplitude  (\ref{e8}).


\section{Conclusions.}  Since the conformal invariance can be broken spontaneously, a dilaton could emerge in the low-energy spectrum.
We have studied the decay of a dilaton into a vector $U$-unparticle and a single photon. For a certain relation between couplings in NCS  the field solutions are defined by 4th order differential equation (\ref{e5}). 
An analytic expression for two-scalar particle correlation function is derived, and the heavy quark interplay due to dilaton field exchange is discussed.
We suggest the dilaton fields are condensed and then the string forms between color (heavy) charges. This is the analog to the Abelian Higgs model.   

A nontrivial  scale invariant sector of dimension $d$ may give rise to peculiar missing
energy distributions in $ \sigma \rightarrow\gamma\,U$ that can be treated in the experiment.
Our results imply that these transitions are near a border of the conformal invariance breaking.
Unless the LHC can collect a very large sample of $\sigma$, the detection of $U$- unparticles
through $ \sigma \rightarrow \gamma\,U$ would be quite challenging. It is related with the results of this work which are useful for many reasons. Among them, in particular, there are:\\
- the couplings of a dilaton are similar to those of the SM Higgs-boson;\\
- a dilaton, if observed, could open the window to the conformal pattern of the strong sector.\\
This would be supported by the study of  $ \sigma \rightarrow \gamma\,U$ where the scale invariant sector is close to EW sector, that could provide the decay $ \sigma \rightarrow \gamma\,\gamma$ to be compared with LHC data in searching for new light scalar object with the mass close to 125 GeV.

\end{document}